\newcommand{\cName}[1]{\texttt{\textbf{#1}}}

\documentclass[
  a4paper,
  10pt, 
  unnumberedsections, 
  twoside, 
]{LTJournalArticle}

\usepackage{listings}
\usepackage{color}

\definecolor{dkgreen}{rgb}{0,0.6,0}
\definecolor{gray}{rgb}{0.5,0.5,0.5}
\definecolor{mauve}{rgb}{0.58,0,0.82}

\runninghead{Distributed Run Manager}

\predate{}
\postdate{}
\date{}

\title{The Distributed Run Manager \\ A C\texttt{++} Library for Distributed Processing} 

\author{%
	Brian P. Harrawood\thanks{Corresponding author: \href{mailto:brian.harrawood@duke.com}{brian.harrawood@duke.com}}
}

\date{\footnotesize Department of Radiology, Duke University - Retired
}

\renewcommand{\maketitlehookd}{
  \begin{abstract}
    \noindent
    The \textbf{DistributedRunManager} (DRM) is a C\texttt{++} library
    for distributing a computing task among a number of computer
    systems for parallel processing on a network. The library was
    originally designed to accelerate synthetic image production for
    tomographic images, but is also a general solution for
    distributing a software task across diverse, multiple hardware
    platforms.
    
    \noindent
    The DRM self-configures using a novel C\texttt{++} class, the
    \textbf{ElectionManager}. This class selects a server
    cooperatively at runtime and eliminates the need for \emph{a
    priori} distribution configuration. All communication, logging,
    and fault-tolerance mechanisms are encapsulated by the DRM,
    providing a ``black-box'' interface to the application
    developer.

    \noindent
    The model for distribution is an unbounded client-server in which
    the clients perform the computing tasks and the server maintains
    the state of the overall progress. The library is tolerant of
    individual processes (clients) joining and/or exiting the work
    pool at run-time, providing dynamic scaling and load balancing.
    
  \end{abstract}
}

\usepackage{bold-extra}
\begin{document}

\maketitle

\section{Introduction}

The advances in High Performance Computing (HPC) have been greatly
accelerated in recent years specifically for computationally intensive
tasks such as scientific modeling. This has led to the development of
more affordable hardware architectures which perform massively
parallel computations. However, a parallel programming model is
required for scientific application software to make use of this
massive parallelism.

Currently, the Message Passing Interface (MPI) standard (see
\url{https://docs.open-mpi.org/}) is used for building distribution
models but requires considerable design and coding to integrate with
an application \autocite{alhafez2025parallelparadigmsmodernhpc}. The
DRM library offers a clear, concise API to the application developer
relieving them of the coding of the distribution architecture, network
communications, fault tolerance and restart capability.

The \cName{g4DistributedRunManager} described in
\autocite{harrawood2014geant4} is an early precursor to the DRM. The
current DRM decouples the code from the \textbf{GEANT4 Toolkit}
\autocite{agostinelli2003geant4} and adds a new class, the
\cName{ElectionManager}. This class permits DRM enabled applications
to self-organize at runtime and select a process from among themselves
to act as a server to manage the state of the overall task.

The DRM is open source released under the GPLv2 license and is
available on GitHub
(\url{https://github.com/brianharrawood/catlib}). User documentation
and sample code are provided there. 

\section{Design Overview}

A distinction is often made between \emph{parallel processing} and
\emph{distributed processing} \autocite{ali2015distributed}. This
difference regularly hinges on whether the run is accomplished using a
single node (i.e., parallel) or multiple nodes connected via a network
(i.e., distributed).  The DRM is designed to use a blend of both
approaches, employing multiple processes on a single node and/or
multiple nodes on a local network.

The DRM scheme for maintaining the state of the run is the
client-server model. In particular, it is the Single Server Multiple
Client System (SSMCS) \autocite{phawade2026verification} in which one
server provides individual parts of the work to multiple clients in
the form of a ``ticket''. The system is unbounded in the number of
clients it can support, and clients are free to come and go as the run
evolves.

The server process keeps track of which tickets are completed, which
are running, and which are ready-to-run. Using message passing, a
client process requests a piece of the run (i.e., a ticket) from the
server. It completes that portion of the run and then reports back to
the server.  A DRM application will also run as a stand-alone process
in which the server asks itself for tickets and processes them
sequentially.

The DRM ``server'' should not be confused with the term server used
for a persistent process such as the \texttt{httpd} server. The DRM
server is created at the beginning of a run and exits when the run is
complete. There are no daemon processes used.

The network bandwidth requirements for communications between clients
and server are modest. Each client must be capable of performing any
piece of the run. Communications between the server and the client is
reduced to just \emph{which} piece to do next and can usually be
represented as an integer or a short string.

\subsection{Designating the server - Election Management}

The client-server algorithm requires that there is one and only one
server. To automate this, the DRM uses the \cName{ElectionManager}
class, which chooses a server from the among running processes. This
runtime election removes the need for a mechanism to do \emph{a
priori} designation of clients and server.

The distinction in the DRM between client and server processes is
handled at runtime with logical operators. The library's single code
base executes in either mode. The DRM process starts out as neither
client nor server. The \cName{ElectionManager} assigns one role or the
other to the process and that role is then immutable.

When a process starts, it broadcasts to the network, seeking a server
process. If a server answers, the process connects as a client. If no
server answers and no other broadcasts are detected, the process
becomes the server.

If there are other processes searching for a server and no server is
responding, an election is held among the processes to select one is
to act as the server. The cooperatively selected process is
``elevated'' to server and the remaining processes attach to it as
clients.

\subsection{Task Granularity - Tickets}

The pieces of the run which are distributed to the clients are encoded
in the C\texttt{++} \cName{Ticket} class. A run is divided into pieces
which can be uniquely mapped to tickets. The amount of work included
in a ticket is sometimes referred to as the granularity of the
distribution. Each ticket completely describes a piece of the run to
be executed within a single client process. If the overall run is
imagined as a logical loop, then each ticket is an iteration of that
loop.

Generally, each ticket uses different input data to generate its own
output data; this is loosely a variation of the Single Instruction
Multiple Data (SIMD) processing algorithm. The output data are written
to file by the client process. At the end of the run, the DRM benefits
from a shared file system (e.g., NFS) to collate these files.

The tickets are dynamically distributed to clients. As a client comes
free, it will be assigned the next ticket by the server. This gives
DRM-enabled applications the flexibility to run with any number of
clients. Additionally, the DRM will accomodate the number of clients
changing \emph{during} the run.

\subsection{Design Constraints}

The DRM is only useful for solving a particular class of problems. The
problems must be ``parallel-izeable'' in nature. In other words, they
must require many repetitive iterations, each of which is independent
of the any other. Iteration $i+n$ may not rely on output from
iteration $i$.

The DRM is an application level tool. It has no facility for starting,
stopping, or reporting on system processes. To effectively use a
DRM-enabled application, these ``meta'' activities must be managed by
another agent: either manually at the command line, or via a Beowulf
cluster \autocite{becker1995beowulf} or using an executive manager
such as slurm (see \url{https://slurm.schedmd.com}).

\section{Implementation}

The DRM is written in C\texttt{++} using the C\texttt{++} Standard
Library and makes extensive use of Inter-Process Communication (IPC)
mechanisms such as sockets, signals and pthreads. These features are
part of the Posix Standard (\url{https://posix.opengroup.org}) and are
available in all conforming operating systems. Library dependencies
include \texttt{libz}, \texttt{libpthread}, and \texttt{libcrypto}, a
part of the OpenSSL Project (\url{https://www.openssl.org/}).

The DRM is implemented with a small number of C\texttt{++} classes:
\cName{DistributedRunManager}, \cName{TicketManager}, \cName{Ticket},
and \cName{ElectionManager}.

\subsection{DistributedRunManager}

A single \cName{DistributedRunManager} object is created per
application to control the distribution. It accepts input from the
application that includes a network port number and a network
key. These values distinguish a ``run,'' and all processes in a run
must agree on their values. The network port number is the port which
the DRM server will use to listen for network connections. The network
key is simply a string passed between processes as a handshake. If the
handshake doesn't agree, communications are ended.

When the \cName{DistributedRunManager} class is first instantiated it
is neither a client nor a server. It creates an
\cName{ElectionManager} to determine which mode it will use going
forward. The \cName{ElectionManager} returns either the address of the
server or informs the process that it is the server. It is here that
the application is committed to run as a server or client until the
run is finished.

\subsubsection{Server initialization}

To communicate with clients, the server starts a dispatcher. This is a
collection of threads used for communication with clients. It starts
with the server creating a thread which listens for new connections on
the network port. This thread \emph{blocks} while waiting for
connections and persists for the life of the
\cName{DistributedRunManager}.

When a client connects, the listen thread creates a new thread
specifically for that client's communications. A unique communication
thread is created for every connected client. These threads
\emph{block} while waiting for client messages and persist only as
long as the client is connected. Client requests for tickets are
answered in these threads.

Meanwhile, the main thread continues, running tickets in the server.

\begin{figure}
	\includegraphics[width=\linewidth]{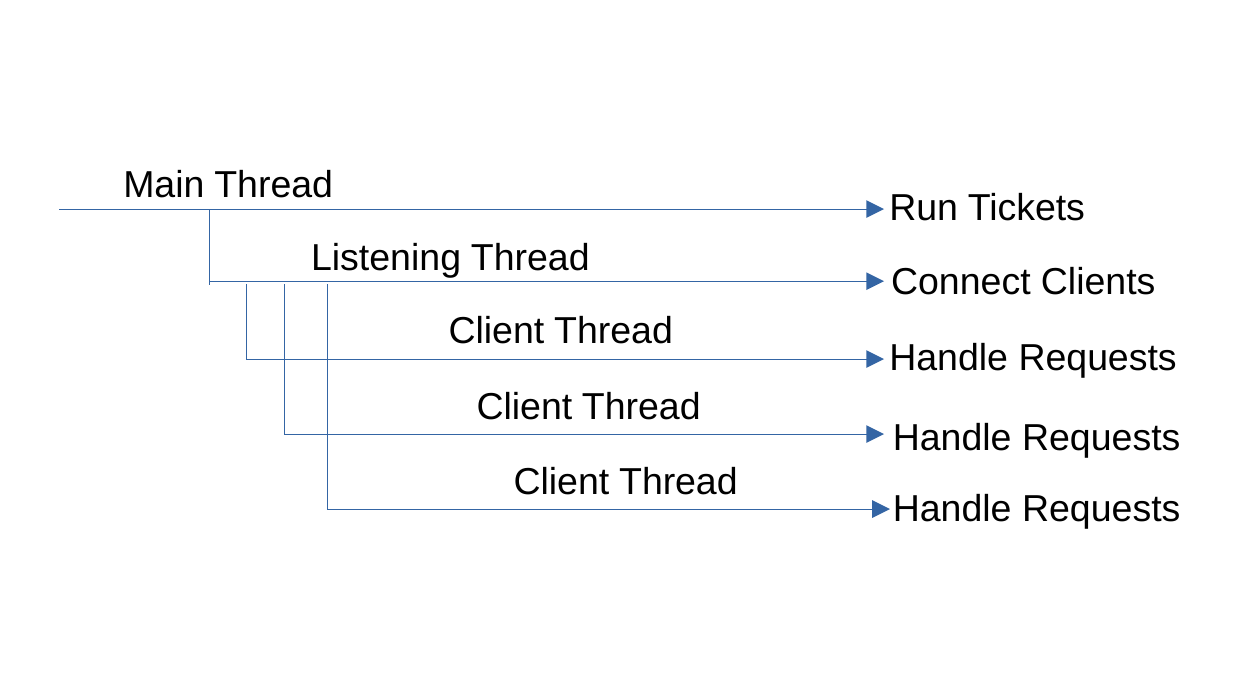}
	\caption{The DRM server process runs multiple threads to act as a
          dispatcher for client requests. The listening thread spawns
          a new thread for each client. The main thread goes on to run tickets.}
	\label{fig:Dispatcher}
\end{figure}

\subsubsection{Client initialization}

In client mode, the DRM connects to the server, exchanges the
handshake, and loads the parameters for the run. It then requests a
ticket and \emph{blocks}, waiting for the server to send work
instructions (i.e., a ticket).

\subsubsection{Network Protocol}

The entire network protocol is comprised of three commands, each
initiated by the client:

\begin{itemize}
  \item Handshake
  \item Parameters
  \item Next Ticket
\end{itemize}

The client's first message to the server is its handshake string.
Part of the handshake string is the \emph{network key}. The key is a
string determined by the application developer. Each application needs
a unique string to prevent DRM-applications from colliding. If the
server and client handshakes disagree, communications are ended.

Next, the client requests the input parameters for the run. Parameters
are input to all the processes from the command line when they are
started. However, significant time may pass between these various
start times. Therefore, the server process provides the input
parameters to each client, guaranteeing consistency.

Finally, the client requests a ticket. When the server responds with a
new ticket, the client performs the work specified by that ticket and
creates any necessary output files. This process repeats until all the
tickets have been exhausted. Then the server returns an
\texttt{EndOfTickets} message and the client exits.
 
\subsection{Tickets}

The \emph{Ticket} is a data representation of a unit of work for the
distribution, a single grain of its granularity. The two C\texttt{++}
classes \cName{Ticket} and \cName{TicketManager} are abstract base
classes from which application specific classes are derived.

\subsubsection{Ticket Class}

The \cName{Ticket} object is primarily a container for the
information the application needs to run a single grain (i.e., an
iteration) of the distribution. Often this is simply an integer.

Each instance has an identification number which corresponds to the
iteration number of the distribution. The ticket ID is passed into the
\cName{Ticket} constructor and a unique ticket is made for each
iteration.

The \cName{Ticket} class also contains two methods for packaging its
information for network transmission as a string:
\texttt{TicketToBuffer()} and \texttt{BufferToTicket()}. Generally,
these simply operate on the ticket ID. This is a \emph{shallow}
transfer of the ticket on the network although \emph{deep} marshaling is
possible.

\subsubsection{TicketManager Class}

The \cName{TicketManager} is designed as an interface between the
\cName{Ticket} class and the \cName{DistributedRunManager} class. It
is created once per application and each DRM must have a registered
\cName{TicketManager} in order to call \texttt{RunTickets()}.

This class has virtual methods for creating a new ticket, running the
code for the ticket, and incrementing the ticket. Creating and
incrementing the ticket are only performed on the server. Both server
and clients, however, run the tickets.

\subsubsection{Running Tickets}

The DRM takes control of the application when a call is made to the
DRM's \texttt{RunTickets()} method. It is called once and it retains
control of the application until all tickets have been processed.

The \texttt{RunTickets()} method works the same in both server and
client modes of the DRM. It simply requests the next ticket and runs
it, until all the tickets have been handled. The request for the next
ticket is always fulfilled by the server, which tracks which tickets
are ready to run. The \texttt{RunTickets()} method returns control to
the application after all tickets have been completed.

The DRM uses a callback function into the application code to execute
a ticket.  This callback function is passed into the
\texttt{TicketManager} constructor, e.g.:

\begin{lstlisting}
    // assign callback function
    catlib::DRMCallbackFunc_t callback_func = run_the_ticket;

    // construct the ticketManager
    MyTicketManager ticketManager(callback_func, &myApp);
\end{lstlisting}

The \texttt{HandleTicket()} method runs this callback with a ticket
argument. The \cName{TicketManager} may unpack information in the
ticket before passing it into the callback function, e.g.:

\begin{lstlisting}
int MyTicketManager::HandleTicket(catlib::Ticket * ticket)
{
  MyTicket * my_ticket = (MyTicket *)ticket;

  int tckNumber = my_ticket->GetTicketNumber();

  CallbackFunc_(tckNumber, CallbackArg_);
  
  return(0);
}
\end{lstlisting}

The callback function, in turn, passes execution to the application. The
application runs the iteration described by the ticket and generates
output.  When the callback returns, execution is passed back to the
DRM. This repeats until the tickets are completed.

\begin{lstlisting}
void run_the_ticket(int ticket, void * arg)
{
  MyApp * my_app = (MyApp *) arg;

  my_app->doCalculation(ticket);
  
  return;
}
\end{lstlisting}

\subsubsection{DRM Fault Tolerance}

The DRM server records each ticket in a file as it is completed. In
the event of a \emph{server} crash, the entire run may be restarted by
enabling the restart option.  This option instructs the DRM to skip all
tickets recorded in the restart file and restart the run where it left
off.

Additionally, whenever the \texttt{RunTickets()} method gets to the
last ticket in the server, the restart file is searched for missing
tickets due to the possible premature exit of a client process from
the distribution. Any missing tickets are (re-)run by the server to
assure completion of the run.

\subsection{ElectionManager}

A DRM distribution must have one and only one server. If only a single
DRM application is started, it will be its own server. If multiple
applications are started, only one of them may be the server. The
remaining applications must become clients.

The \cName{ElectionManager} cooperates in a pool of other
applications using the DRM to select a server for the distribution. It
is not called directly by the application. Instead a single instance
is created by the \cName{DistributedRunManager} when it starts.

When the \cName{ElectionManager} starts, it attempts to open a network
port on the local machine.  Only one process may bind to a network
port on a host. If multiple \cName{ElectionManager}s are started
\emph{on the same host}, only one of them (the first) will be able to
open the network port. This is the \emph{Network} instance.

Subsequent calls to bind that port return an error. Although these
instances cannot open the network port they still run, but they run as
\emph{Local} instances. They connect to the \emph{Network} instance on
their host, but are not candidates in the network election.
 
\subsubsection{Network ElectionManager}

If opening the network listening socket is successful (i.e., the
\emph{Network} instance), the \cName{ElectionManager} broadcasts
inquiry messages to the network repeatedly for \texttt{TMOUT\_INQUIRY}
seconds. If a server is available, it responds with its IP
address. This information is passed back to the DRM, which will then
connect as a client to that address.

If no server answers during \texttt{TMOUT\_INQUIRY} seconds
\emph{and} no inquiries arrive from other processes also seeking a
server, then this process becomes the server.

However, if no server answers after \texttt{TMOUT\_INQUIRY} seconds,
and there \emph{are} other processes seeking the server, then an election is
started. Those other processes seeking a server form a candidate list.

The candidate list is sorted by IP address. All IP addresses greater
than ``mine'' are put into a nominee list, and a nomination message is
sent to each nominee. Meanwhile, every nomination message received is
answered with a nomination accepted message.

After \texttt{TMOUT\_ELECTION} seconds, the list of hosts accepting
their nomination is examined. On one host, the list will be empty, and
that process becomes the server, i.e, it answers all pending and
new broadcast messages seeking the server with its own address. If the
list is not empty, an inquiry broadcast is repeated until the elected
server reports back.

\subsubsection{Local ElectionManager}

Processes without access to the network port (i.e., the \emph{Local}
instances) open a \emph{unix} socket with which they can communicate
with the \emph{Network} process running on their host machine. When
the \emph{Network} process has determined the server's IP address, it
informs all \emph{Local} processes, which may then connect to the
network port of the server and act as clients there.

After a server is selected, its \cName{ElectionManager} responds to
new broadcasts on the election port with its IP address, and all
processes starting after the election are directed to the same
server.

\begin{figure}
	\includegraphics[width=\linewidth]{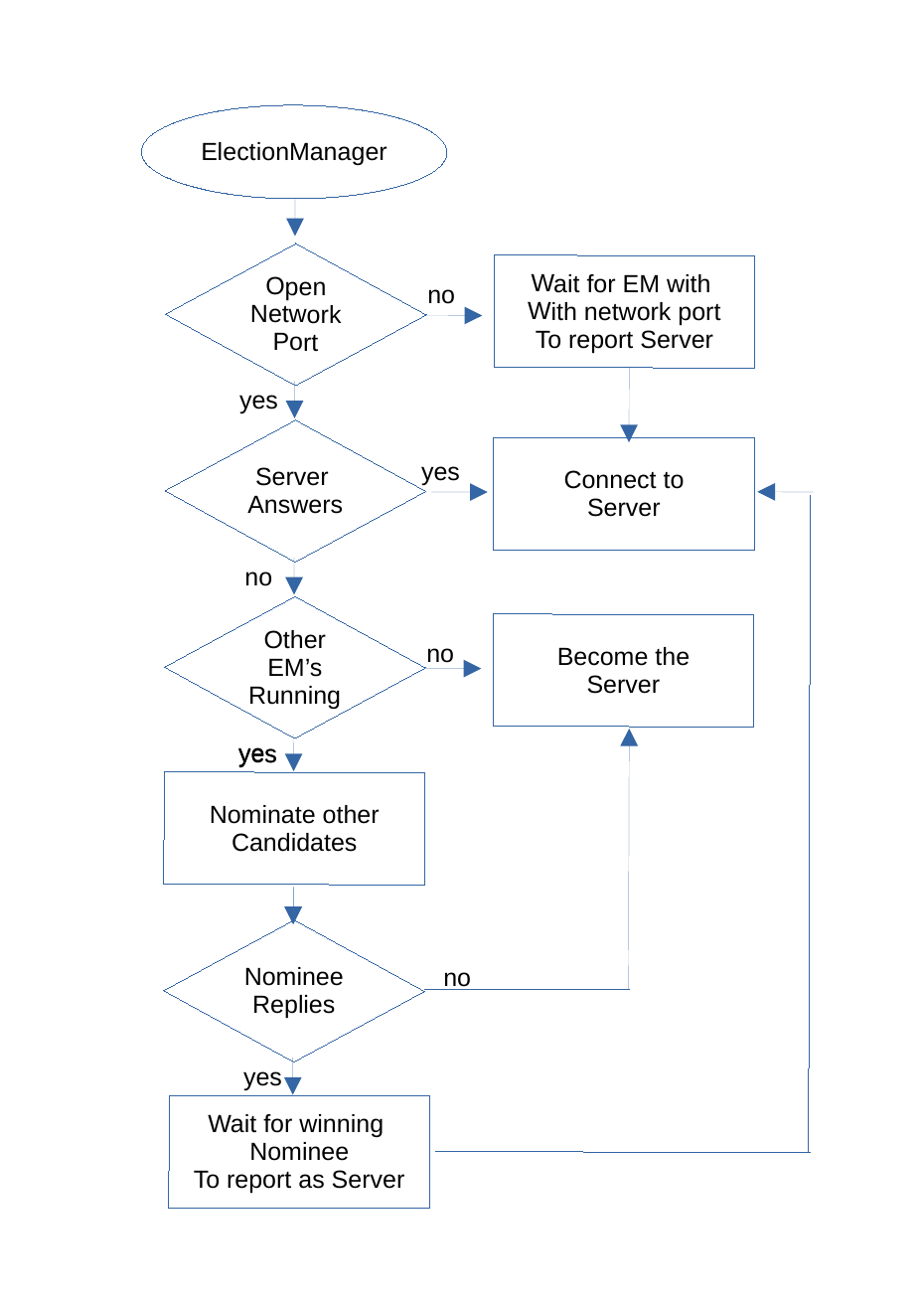}
	\caption{The DRM's ElectionManager negotiates asynchronously with other
          ElectionManagers running on the host and/or network to
          determine which will become the server.}
	\label{fig:ElectionManager}
\end{figure}

\section{Conclusion}

Much effort has been devoted to distributed computing
\autocite{ali2015distributed} with the advent of less expensive
computers and more reliable network connections. However, applications
must be modified to leverage the efficiencies of these distribution
systems. The DRM offers a reusable tool which provides a single
interface to the distribution problem.

The DRM interface hides the message passing, client-server selection
process, client enrollment, and error handling of the distribution. It
scales without modification from a single process to a High Performance
Compute Cluster.

The DRM has been used for many years, for varied application types,
and in many environments. It is a robust and
flexible solution for transforming a serial application into a
distributed application.

\section{Acknowledgments}

The DRM was first begun in the 2003-2004 time frame as a solution to
increase throughput of Monte Carlo analyses, specifically for the
NSECT program \autocite{floyd2004neutron}. Its steady development over
the 20-plus years since then was made possible by Duke University's
Radiology Department.

The author is grateful for the support and patience of the many
researchers he worked with during these decades. Without the support
of Rai Labs Advanced Imaging (\url{https://rai.labs.duke.edu}), and the
Center for Virtual Imaging Trials (\url{https://cvit.duke.edu}), this
work would never have been completed.

\printbibliography

\end{document}